\documentclass[english,12pt]{article}
\usepackage{array}
\usepackage{graphicx}
\usepackage{amssymb}
\usepackage[english]{babel}
\usepackage{amsmath}
\usepackage{multirow}
\usepackage{prettyref}
\usepackage{babel}
\usepackage{units}
\usepackage[latin1]{inputenc}
\usepackage{amsfonts}
\usepackage{amssymb}
\usepackage{babel}
\usepackage{color}
\usepackage{cite}
\def\@fmsl@sh#1#2#3{\m@th\ooalign{$\hfil#1\mkern#2/\hfil$\crcr$#1#3$}}
 \def\eq#1\en{\begin{equation}#1\end{equation}}
\def\s[#1,#2]{[#1\stackrel{\star}{,}#2]}
\def\sx[#1,#2]{[#1\stackrel{\star_{x}}{,}#2]}

\newcommand{\nc}{\newcommand}
\nc{\beq}{\begin{equation}}
\nc{\eeq}{\end{equation}}
\nc{\beqa}{\begin{eqnarray}}
\nc{\eeqa}{\end{eqnarray}}

\def\bc{\begin{center}}
\def\ec{\end{center}}

\def\gsim{\mathrel{\mathpalette\atversim>}}

\def\bc{\begin{center}}
\def\ec{\end{center}}

\def\gsim{\mathrel{\rlap{\lower4pt\hbox{\hskip1pt$\sim$}}

    \raise1pt\hbox{$>$}}}       

\def\gsim{\mathrel{\rlap{\lower4pt\hbox{\hskip1pt$\sim$}}
    \raise1pt\hbox{$>$}}}       

\begin{document}
\makeatletter
\def\fmslash{\@ifnextchar[{\fmsl@sh}{\fmsl@sh[0mu]}}
\def\fmsl@sh[#1]#2{%
  \mathchoice
    {\@fmsl@sh\displaystyle{#1}{#2}}%
    {\@fmsl@sh\textstyle{#1}{#2}}%
    {\@fmsl@sh\scriptstyle{#1}{#2}}%
    {\@fmsl@sh\scriptscriptstyle{#1}{#2}}}
\def\@fmsl@sh#1#2#3{\m@th\ooalign{$\hfil#1\mkern#2/\hfil$\crcr$#1#3$}}
\makeatother

\thispagestyle{empty}
\begin{titlepage}
\boldmath
\begin{center}
  \Large {\bf  Cosmological Time Evolution of the Higgs Mass \\ and \\ Gravitational Waves}
    \end{center}
\unboldmath
\vspace{0.2cm}
\begin{center}
{  {\large Xavier Calmet}\footnote{x.calmet@sussex.ac.uk}}
 \end{center}
\begin{center}
{\sl Department of Physics and Astronomy, 
University of Sussex, Brighton, BN1 9QH, United Kingdom
}\end{center}
\vspace{5cm}
\begin{abstract}
\noindent
We point out that gravitational wave detectors such as LISA have the potential of probing a cosmological time evolution of the Higgs boson self-coupling constant $\lambda$ and thus the Higgs boson's mass $m_H= \sqrt{ 2 \lambda} v$. The phase transition of the Standard Model could have been a first order one if the Higgs mass  was below 72 GeV at a temperature $T_\star \ge 100$ GeV. Gravitational waves could thus have been produced during the electroweak phase transition. A discovery by LISA of a stochastic background of gravitational waves with a characteristic frequency $k_\star \ge 10^{-5}$ Hz could be interpreted as a sign that the Higgs boson self-coupling constant was smaller in the past. This interpretation would be particularly tempting if the Large Hadron Collider did not discover any physics beyond the Standard Model by the time such waves are seen. The same mechanism could also account for baryogenesis.
\end{abstract}  
\vspace{5cm}
\end{titlepage}



\newpage

\section{Introduction}	
In this paper we point out that future gravitational wave detectors have the potential of probing a cosmological time evolution of the parameters of the Higgs boson potential given by $V=\mu^2 H^\dagger H+\lambda (H^\dagger H)^2$, where $H$ is the Higgs doublet. The Higgs particle \cite{Higgs:1964ia,Englert:1964et,Higgs:1964pj} that was discovered at CERN in 2012 is one of the simplest yet  most surprising objects in the Standard Model for several reasons. It is the only fundamental scalar field that we have discovered thus far. It mediates a new force between the particles of the model, yet this force is not mediated by a gauge boson in stark contrast to other fundamental forces of nature. The Higgs particle generates masses for all fields of the Standard Model and it is the source of the spontaneous breaking of the  $SU(2)_L\times U(1)_Y$ local gauge symmetry. Yet very little is known about the Higgs field besides the fact that in our cosmological era, its mass is 125 GeV. The Higgs field could have played a very important role in the history of our universe, potentially being at the origin of inflation \cite{Bezrukov:2007ep,Barvinsky:2008ia,Barvinsky:2009fy}. It is thus natural to ask whether the parameters of the Higgs sector, i.e. the mass of the Higgs boson or its self-coupling could have had a cosmological time evolution \cite{Calmet:2017czo}. At this point it is worth emphasizing that there are very well motivated models which can lead to a cosmological evolutions of fundamental constants, see e.g. Ref. \citen{Uzan:2010pm} for a recent review.  These models range from Kaluza-Klein theories \cite{Marciano:1983wy} and other string inspired models to Bekenstein's models \cite{Bekenstein:1982eu,Bekenstein:2002wz}, chameleons models \cite{Khoury:2003aq} and  quintessence models \cite{Dvali:2001dd}.

\section{Time Variation in the Higgs Sector}
	
In the Standard Model, the Higgs boson's mass is fixed by the following relation between the Higgs vacuum expectation value $v=246$ GeV and the Higgs boson self-coupling $\lambda$:
\begin{equation}
    m_H= \sqrt{-2 \mu^2}=\sqrt{2 \lambda} v.
\end{equation}
The vacuum expectation of the Higgs boson fixes the masses of the electroweak bosons and thus Fermi's constant which itself determines the strength of the electroweak interactions. The Higgs boson's mass could have a cosmological time evolution if its self-coupling, the vacuum expectation value or both of them had a cosmological time evolution. 

Nucleosynthesis tells us that Fermi's constant must have been within 10 to 20$\%$ of its present value at $T=1$ MeV \cite{Scherrer:1992na}. On the other hand, there is virtually no information about the value of the Higgs boson self-coupling. Thus, a smaller Higgs boson's mass at the electroweak phase transition time or some $T=100$ GeV could easily be obtained by a cosmological time evolution of $\lambda$. Note that the parameter of the potential $\mu^2$ merely sets the temperature of the phase transition, not its strength which is determined by $\lambda/g^2$ where $g$ is the gauge coupling of the weak interactions. The only constraint on a cosmological time evolution of $\lambda$, is that the physical Higgs boson's mass $m_H$ must reach its present value by today's era. Furthermore, the Higgs boson should not have been lighter than 3.72 GeV \cite{Weinberg:1976pe,Linde:1975sw} in the early universe otherwise the early universe vacuum would have been unstable. We note that a similar effect could be obtained by a cosmological time evolution of the top quark mass which plays an important role in determining the strength of the electroweak phase transition as it impacts the shape of the thermal Higgs potential.

Future gravitational wave detectors such as LISA have the potential of probing a cosmological time evolution of the Higgs mass, or rather to be precise, of the self-coupling parameter $\lambda$. Indeed if the Higgs mass was smaller than 72 GeV, because its self-coupling was smaller in the past, the electroweak phase transition in the Standard Model would have been a first order one without the need for any new physics beyond the Standard Model at this scale \cite{Kajantie:1996mn,Rummukainen:1998as,Csikor:1998eu}. While in the cosmological time-independent Standard Model with a Higgs boson of 125 GeV, the phase transition is not a first order one and thus do not lead to the production of gravitational waves, in a cosmological time dependent Standard Model, the universe may have undergone a first order phase transition if the Higgs boson's mass was below 72 GeV at a temperature $T_\star \ge 100$ GeV. The resulting stochastic gravitational wave background would have a characteristic frequency $k_\star \ge 10^{-5}$ Hz and could be observable by LISA (see e.g. Refs. \citen{Caprini:2010kz,Caprini:2015zlo}, for useful reviews). If such a signal is seen by LISA,  it could be interpreted as a sign that the Higgs mass was smaller in the past. 

One may worry that the electroweak vacuum stability is a serious constraint for the mechanism described here. 
Indeed, in the Standard Model of particle physics, quantum corrections affect the shape of the the Higgs potential which is taken at the classical level to have its usual sombrero potential shape. The Higgs boson is assumed to settle at the minimum of this potential.  However, at the quantum level, the Higgs potential develops a second minimum besides the one at which the Higgs field takes its usual vacuum expectation value. The location and depth of this second minimum mainly depend on the Higgs boson and top quark masses, $M_H$ and $M_t$, and for the known values, $M_H = 125$ GeV and $M_t = 173.34$ GeV, it turns out to be much deeper than the usual electroweak one, thus being a false vacuum (to be precise a metastable state). The universe, while it may be sitting in the usual electroweak vacuum is expected to decay to the true lower vacuum with potentially catastrophic consequences. 

A light Higgs boson's mass implies that the electroweak vacuum was unstable in the past, while it might be metastable today. However, while it is clear that the electroweak vacuum would be unstable if the Higgs mass was below 72 GeV at zero temperature, one should not forget that temperature dependent corrections will stabilize the electroweak vacuum even for such a light Higgs boson. The only requirement is thus that shortly after the phase transition the Higgs boson self-coupling reaches today's value fast.

\section{Baryon Asymmetry}

A cosmological time evolution of the Higgs boson's mass could also help to explain the baryon asymmetry in our universe. Electroweak baryogenesis is a very elegant mechanism which unfortunately does not work in the Standard Model of particle physics as the amount of CP violation is too weak to lead to the observed baryon asymmetry and because the phase transition is not first order for a Higgs mass of 125 GeV. As we have just argued, the second problem can be solved by a cosmological time dependence of the Higgs mass via its self-coupling.

 The issue linked to the amount of CP violation could be solved by a similar mechanism. Indeed, it has been shown in Ref. \citen{Berkooz:2004kx}, that if the Yukawa couplings of the Standard Model were of order one in the early universe, i.e. if they had a cosmological time evolution, then the CP phase present in the CKM matrix could be large enough to explain the observed baryon asymmetry. We note that the authors of Ref. \citen{Berkooz:2004kx} also considered a cosmological time evolution of the Higgs boson's mass  triggered by a time variation of the vacuum expectation of a singlet scalar field in the context of the Froggatt-Nielsen Mechanism \cite{Froggatt:1978nt}. Small Yukawa couplings and their hierarchy can indeed be explained by that mechanism which invokes a horizontal Abelian symmetry. In that model, fields of different generations carry different charges under this horizontal symmetry. This symmetry is spontaneously broken by a vacuum expectation value of a scalar field which is a singlet under the gauge group of the Standard Model. The symmetry breaking is communicated to the fields of the Standard Model via  quarks and leptons in vector representations of the Standard Model gauge group. It is plausible that the vacuum expectation value of scalar fields can be cosmological time dependent. If the vacuum expectation value of this new singlet scalar field is time cosmological dependent, then the Yukawa couplings could also have cosmological time evolution as well. 

For sphaleron processes not to wash out the baryon asymmetry, the Higgs boson's mass would have to be low enough to suppress sphaleron processes in the broken phase. The Higgs boson's mass has to be smaller than some $35$ GeV \cite{Grojean:2004xa} for sphaleron processes to be suppressed. The Standard Model could account for the baryon asymmetry without the need for new particles if the Yukawa couplings and the mass of the Higgs boson reach their observed values at the present era before nucleosynthesis. 

If searches at the Large Hadron Collider for physics beyond the Standard Model remain unfruitful, then a cosmological time evolution of fundamental constants should be considered as a leading contender to explain the baryon asymmetry of the universe. We have identified a clear signature in terms of a stochastic background of gravitational waves which could be observed with LISA. It might however be difficult to differentiate this model from models with a phase transition in a hidden sector, see e.g. Refs. \citen{Schwaller:2015tja,Fairbairn:2019xog}.

\section{Conclusions}

In this short paper, we have pointed out that future gravitational wave detectors such as LISA have the potential of probing a cosmological time evolution of the Higgs boson self-coupling constant $\lambda$ and thus the Higgs boson's mass $m_H= \sqrt{ 2 \lambda} v$. The phase transition of the Standard Model could have been a first order one if the Higgs mass  was below 72 GeV at a temperature $T_\star \ge 100$ GeV. Gravitational waves could thus have been produced during the electroweak phase transition. A discovery by LISA of a stochastic background of gravitational waves with a characteristic frequency $k_\star \ge 10^{-5}$ Hz could be interpreted as a sign that the Higgs boson self-coupling constant was smaller in the past. This interpretation would be particularly tempting if the Large Hadron Collider did not discover any physics beyond the Standard Model by the time such waves are seen. The same mechanism could also account for baryogenesis.

A cosmological time evolution of the Higgs boson self-coupling could be explained by a variety of models. In particular, it could be a sign of models of quantum gravity with compactified extra-dimensions such as Kaluza-Klein models. In these models, dimensionless fundamental constants or their ratios are related to the radius of these extra-dimensions which could be contracting, expanding or oscillating during the lifetime of our universe. A discovery of a cosmological time variation of fundamental constants could thus be the first observable signature of quantum gravity.

\section*{Acknowledgments}

This work supported in part  by the Science and Technology Facilities Council (grant number ST/P000819/1).

\end{document}